# Identifying multi-scale communities in networks by asymptotic surprise

Ju Xiang[1,2,3], Yan Zhang[1,2,3], Jian-Ming Li[1,2], Hui-Jia Li[4], Min Li[1,*]

[1]School of Information Science and Engineering, Central South University, Changsha 410083, China.

[2]Neuroscience Research Center & Department of Basic Medical Sciences, Changsha Medical University, Changsha, 410219, Hunan, China.

[3]Department of Computer Science, Changsha Medical University, Changsha, 410219, Hunan, China.

[4]School of Management Science and Engineering, Central University of Finance and Economics, Beijing 100080, China

* Corresponding authors: Min Li.

E-mail: limin@mail.csu.edu.cn (ML)

**Abstract**

Optimizing statistical measures for community structure is one of the most popular strategies for community detection, but many of them lack the flexibility of resolution and thus are incompatible with multi-scale communities of networks. Here, we further studied a statistical measure of interest for community detection, asymptotic surprise, an asymptotic approximation of surprise. We discussed the critical behaviors of asymptotic surprise in phase transition of community partition theoretically. Then, according to the theoretical analysis, a multi-resolution method based on asymptotic surprise was introduced, which provides an alternative approach to study multi-scale communities in networks, and an improved Louvain algorithm was proposed to optimize the asymptotic surprise more effectively. By a series of experimental tests in various networks, we validated the critical behaviors of the asymptotic surprise further and the effectiveness of the improved Louvain algorithm, displayed its ability to solve the first-type resolution limit and stronger tolerance against the second-type resolution limit, and confirmed its effectiveness of revealing multi-scale community structures in multi-scale networks.

# Contents

## 1. Introduction

Complex networks provides a kind of useful way to the study of complex systems, e.g., the metabolic networks and protein-protein interaction networks, and it was revealed that the networks possess many common topological properties [1]. For example, community structure or modular structure have been found to exist widely in various complex networks, meaning the networks consist of groups of densely connected vertices that are sparsely connected with the rest of the networks. The community structure is of interest for understanding the structures and functions of the networks as well as the dynamics on the networks [2-6]. For instance, it was found that local targeted immunization outperforms global targeted immunization in the network with apparent community structure [7]; the abundance of communities in social networks can foster the formation of cooperation under strong selection [8]. Therefore, community detection in complex networks attracted much attention from various fields.

    Many methods have been proposed to identify the communities in complex networks by various approaches [9-18], such as spectral analysis [18], random walk [19-21], dynamics [22-25], label propagation [26], and modularity optimization [27, 28]. The existing methods could indeed help reveal intrinsic structures in the networks, but they also have respective scopes of application, and thus it is necessary to study their behaviors, e.g., the resolution in community detection [29-35]. This could help understand the methods themselves in depth and promote the development of community-detection methods. For example, methods based on modularity optimization and Bayesian inference were found that there exist phase transitions from detectable to undetectable structures in community detection, which provides a bound on the achievable performance of the methods [29-31]. Botta et al presented a detailed analysis of modularity density, showing its superiors and drawbacks [32]. The

original modularity was found to be unable to identify community structure below a certain characteristic scale especially in large networks, known as the (first-type) resolution limit [33], and many other quality functions have similar phenomena. Various approaches have been used to improve the modularity-based methods [11, 36, 37]. Lai et al proposed the improved modularity-based method by random walk network preprocessing [37], and then enhanced the modularity-based belief propagation method by using the correlation between communities to improve the estimate of number of communities [11].

The resolution limit also means that the networks may possess community structures at multiple scales [1], and suggests that it is necessary to develop community-detection algorithms with tunable resolution. In recent years, various multi-resolution methods have been proposed to study the multi-scale community structures in complex networks [10, 38-42]. Some methods make use of the correlation between dynamics and multi-scale structures in networks [21, 43]. Some methods make use of the local optimization of fitness functions [44]. Some methods make use of Potts spin model [26, 41, 45-47]. Especially, it is one of the most effective ways to the resolution limit to introduce a tunable resolution parameter into such quality functions as the modularity [38, 42]. Recently, we proposed one uniform framework for the multi-resolution modularity methods based on the general rescaling strategy [10]. Many important quality functions can be unified in the framework [41, 42, 47], while each of Hamiltonian based on Potts model can also find its counterpart of modularity (corresponding to the negative of the corresponding modularity).

Optimizing statistical measures for community structures is one of the most popular methods for community detection, such as modularity [48], Hamiltonians [41], Partition density [49, 50]. In literature, Aldecoa et al proposed a statistical measure of interest for community structure, (original) surprise. It is defined as the minus logarithm of the probability that the observed number of intra-community links or more is found in Erdös-Rényi random networks [51]. According to a cumulative hyper-geometric distribution, it can be written as,

$$S = -\ln \sum_{j=m_{\text{int}}}^{\min(m, M_{\text{int}})} \frac{\binom{M_{\text{int}}}{j}\binom{M - M_{\text{int}}}{m - j}}{\binom{M}{m}}, \qquad (1)$$

where $M$ is the maximal possible number of links in a network; $M_{\text{int}}$ is the maximal possible number of intra-community links in a given partition; $m$ is the number of existing links in the network; while $m_{\text{int}}$ is the number of existing intra-community links in the partition. It exhibited good performance in many networks [4, 51], but it was proposed originally for un-weighted networks and it involves complex nonlinear factors, leading to the difficulties of the theoretical analysis and numerical computations. Recently, Traag et al [52] proposed a kind of accurate asymptotic approximation for surprise, called asymptotic surprise (AS), while we call the surprise of Aldecoa et al as original surprise (OS) to avoid confusion. By only taking into account the dominant term and using Stirling's approximation of the binomial coefficient, the asymptotic surprise reads,

$$\begin{aligned} S &\approx m\left( q\log\frac{q}{\bar{q}} + (1-q)\log\frac{1-q}{1-\bar{q}} \right), \\ &= mD(q \parallel \bar{q}) \end{aligned} \qquad (2)$$

where $q = m_{int}/m$ denotes the probability that a link exists within a community; $\bar{q} = M_{int}/M$ denote the expected value of $q$; $D(x\|y) = x\ln\frac{x}{y} + (1-x)\ln\frac{1-x}{1-y}$ is the Kullback-Leibler divergence, which measures the distance between two probability distributions $x$ and $y$. The asymptotic expression of surprise makes surprise be extended to weighted networks naturally and is helpful for the theoretical analysis for the measure, but it is still a single-scale method with limited resolution.

In this paper, we further discuss the critical behaviors and resolution limit of the asymptotic surprise in phase transition of community partition. The original asymptotic surprise closely depends on the difference between the probability of links existing within communities and the expected values in the random model, so, by using a resolution parameter to adjust the random model, a multi-resolution method based on asymptotic surprise is introduced naturally, which is an extension of asymptotic surprise to multi-scale networks. To optimize the asymptotic surprise more effectively, we propose an improved Louvain algorithm. Then, we conduct a series of experimental tests in various networks to respectively validate the critical behaviors of the asymptotic surprise further and the effectiveness of the improved Louvain algorithm, show the ability of the multi-resolution asymptotic surprise to solve the first-type resolution limit and stronger tolerance against the second-type resolution limit, and confirm its effectiveness of revealing multi-scale structures in a set of homogeneous networks and a set of heterogeneous networks. Lastly, we come to conclusion.

## 2. Methods

To provide a theoretical basis for the extension of asymptotic surprise, we firstly discuss the critical behaviors and resolution limit of it, by analytically deriving the critical number of communities in community merging, and then introduce a multi-resolution method based on asymptotic surprise and an improved Louvain algorithm for optimizing asymptotic surprise.

### 2.1. Critical behavior of asymptotic surprise and its resolution

For convenience of analysis, we introduced a set of community-loop networks with $r$ communities that are connected one by one (see Appendix). To display the critical behaviors of asymptotic surprise in partition transition, we consider a set of partitions that consists of $r/x$ groups of vertices, where each group contains $x$ adjacent communities. For the partitions, the asymptotic surprise can be written as,

$$S_x = m\left(q_x \ln\left(\frac{q_x}{\bar{q}_x}\right) + (1-q_x)\ln\left(\frac{1-q_x}{1-\bar{q}_x}\right)\right)$$
$$= m\left(\left(1-\frac{2\xi}{1+2\xi}\frac{1}{x}\right)\ln\left(\left(1-\frac{2\xi}{1+2\xi}\frac{1}{x}\right)\bigg/\frac{x}{r}\right) + \frac{2\xi}{1+2\xi}\frac{1}{x}\ln\left(\frac{2\xi}{1+2\xi}\frac{1}{x}\bigg/\left(1-\frac{x}{r}\right)\right)\right) \quad (3)$$

where $\xi = p_o/p_i$, $q_x = 1 - \frac{2\xi}{1+2\xi}\frac{1}{x}$ and $\bar{q}_x = x/r$ (see Appendix). In the networks, the pre-defined partition is a special case of the partition with $x=1$, when $x \geq 2$ communities will be merged.

The asymptotic surprise, as a multivariate function, is closely related to various network parameters. **Figure 1**(A) shows that, for small $r$-values, $S(x)/S(1)$ decreases with $x$, and $S(x)/S(1)<1$, that is to say, $S(x)<S(1)$. This means there is no appearance of community merging. For large *r*-values, there is a peak where $S(x)/S(1)>1$, meaning the appearance of community merging. **Figure 1**(B) shows that, $S(r)$ increases with the increase of *r*. By comparing $S(r)$-curves of different *x*-values, for small *r*-values, $S(r,x)/S(r,1)<1$, i.e. $S(r,x) < S(r,1)$, meaning there is no merging of communities; with the increase of *r*, $S(r, x=2)$ and $S(r, x=3)$ will be larger than others in turn, meaning the community merging for *x*=2 and 3 will be preferred. **Figure 1**(C) shows, with the increase of $\xi = p_o/p_i$, the (normalized) *S*-curves decrease for different *x*-values, and $S(x=1, 2$ or $3)$ will be larger than others in turn. This means that the partition for *x*=1, 2 and 3 will be preferred in turn. Other statistical measures such original surprise and modularity have similar behaviors, but the critical points are different for different statistical measures [34].

To further study the critical points of asymptotic surprise for community merging, consider the transition of partition from *x*=1 to 2 (see Appendix). Community merging will occur when $\Delta S = (S_2 - S_1)/m > 0$. By using $r-1 \approx r-2 \approx r$ for large *r*-value,

$$\Delta S \approx -D(q_1 \| q_2) - q_1 \ln 2 + \beta \ln\left(\frac{q_2}{1-q_2}\frac{r}{2}\right), \qquad (4)$$

where $\beta = \xi / (1+2\xi)$. By solving $\Delta S = 0$ for *r*, we obtain the critical number of communities,

$$\begin{aligned} r^* &\approx \frac{1-q_2}{q_2} \exp\left(\frac{1}{\beta} D(q_1 \| q_2)\right) x^{\frac{q_1}{\beta}+1} \\ &= \frac{\xi}{1+\xi}\left(\frac{1}{1+\xi}\right)^{\frac{1}{\xi}} 2^{\frac{1}{\xi}+3} \end{aligned}. \qquad (5)$$

For comparison, the critical point of modularity for the partition transition in the networks is given by $r^* = 2 + 1/\xi$. Compared to modularity, the critical number of asymptotic surprise has strong nonlinear effect.

**Figure 1**(D) shows a phase diagram where the community-merging partition occurs in the region above the corresponding curve, meaning the existence of resolution limit, while not in the region below the curve. For comparison, we also display the critical points of other measures (original surprise and modularity) in the networks. The resolution of asymptotic surprise decreases with the increase of $\xi$ (i.e. $p_o/p_i$), and so do other measures. This is because the number of links between communities increases and thus the community structures become more and more unclear. By comparing the measures, the asymptotic surprise has higher resolution than modularity, while it is lower than the original surprise.

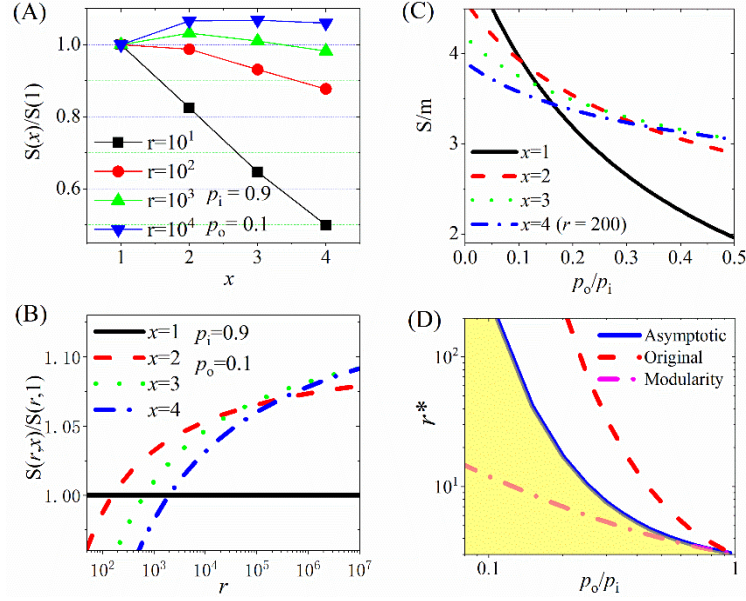

**Figure 1.** (A) In the different-size networks, asymptotic surprise as a function of the number $x$ of merged communities, normalized by $S(x=1)$, i.e. the asymptotic surprise of the pre-defined partition. (B) Asymptotic surprise for distinct $x$-values as a function of the number $r$ of pre-defined communities, normalized by the $S$-values of the pre-defined partition. (C) Asymptotic surprise as a function of $p_o/p_i$ for distinct $x$-values, normalized by the number $m$ of edges in the networks. (D) Phase diagram in partition transition shows critical number of communities in community merging as a function of $p_o/p_i$, for asymptotic and original surprise and modularity.

**2.2. Multi-resolution method based on asymptotic surprise**

Extending asymptotic surprise to multi-scale case is very necessary, because it has only limited resolution and multi-scale structures extensively exist. Before constructing the multi-resolution method based on asymptotic surprise, we firstly recall the definition of the multi-resolution modularity. The original modularity is defined as the fraction of edges within communities in a network minus the expected value in a random graph (i.e. a null model), and the larger modularity generally means the better division. To extend modularity to multi-scale case, the simplest and effective way is to introduce a tunable resolution parameter to adjust the weight of the null model.

Similarly, the original asymptotic surprise is based on the difference between the probability of links existing within communities and the expected values in a random model (also call null model). So, similarly to the multi-resolution modularity, we introduce a multi-resolution method based on asymptotic surprise, by using a resolution parameter to adjust the expected values in the random model. It can be written as,

$$S(\gamma) = m\left( q\log\frac{q}{\tilde{q}} + (1-q)\log\frac{1-q}{1-\tilde{q}} \right), \qquad (6)$$

where $\tilde{q} = \gamma \cdot \bar{q}$ and $\gamma$ is the resolution parameter.

As a result, the critical point of asymptotic surprise for community merging can be rewritten as,

$$r^* \approx \gamma \frac{\xi}{1+\xi}\left(\frac{1}{1+\xi}\right)^{\frac{1}{\xi}} 2^{\frac{1}{\xi}+3}, \qquad (7)$$

Similarly, the critical point of the multi-resolution modularity can be rewritten as $r^* = \gamma(2+1/\xi)$. This suggests that the resolution will increase with the increase of the resolution parameter. By adjusting the resolution parameter, the multi-resolution asymptotic surprise can help identify the communities that are undetectable for the original one as well as the community structures at different scales.

### 2.3. General procedure for optimizing asymptotic surprise

Like modularity, community structures in networks can be identified by optimizing asymptotic surprise. In principle, any suitable optimization algorithms may be used. Here, asymptotic surprise is optimized by the Louvain procedure, which is a fast and efficient way for modularity optimization [28]. However, the strong nonlinearity of asymptotic surprise makes it more difficult to be optimized. To optimize asymptotic surprise more effectively, we therefore introduced two strategies to improve original Louvain procedure. The general procedure for the improved Louvain algorithm is as follow.

(1) Firstly, assign each vertex into a sole group index and calculate the number of common neighbors (CN) between ends of each existing edge.
(2) Randomize the order of the list of vertices and move each vertex into the group that its neighbor with maximal CN belongs to.
(3) Repeat from step (2) 1-2 times to generate a pre-condensation of vertices for community structure.
(4) Select a vertex randomly and move it into the group that generates maximal increment of asymptotic surprise.
(5) Repeat from step (4), until there is no improvement, or improvement reaches predefined value.
(6) Transform current network into a super network, where each group of vertices in current division is considered as a super vertex, the number of links between groups is considered as the weight between super vertices, and the links within groups are considered as the self-loop of super vertices.
(7) Repeat from step (4), until no improvement can be obtained, or improvement reaches predefined value.
(8) Recover the community division of final super-network above into the community division of original network.
(9) Select a vertex (of original network) randomly and move it into the group that generates maximal increment of asymptotic surprise.
(10) Repeat from step (9) until there is no improvement.

Original Louvain algorithm (OL) needs an initial division, while the division often is given by assigning each vertex into a single-vertex group. In fact, it can find the optimal division more effectively, if a better initial division (which is more near to optimal division) is given. Therefore, our first strategy (step (1)-(3)) is proposed to improve the initial division, and the second strategy (step (8)-(10)) is used to further refine the division. As expected, the improved Louvain algorithm (IL) can indeed better find the community structure in networks (see **Figure 2** and **Figure 4** for the comparison between IL and OL algorithms). For the sake of brevity, AS-OL and AS-IL are used to denote the asymptotic surprise that is optimized respectively by OL and IL algorithms, while OS-OL and OS-IL are used to denote the original surprise that is optimized respectively by OL and IL algorithms.

## 3. Experimental results

In this section, firstly we experimentally exhibit the limited resolution of the single-scale asymptotic surprise and give a comparison with other measures, in the above loop-community networks and the Lancichinetti-Fortunato-Rachicchi (LFR) networks (a kind of test networks with more realistic network properties) [53]. Secondly, we exhibit how the multi-resolution asymptotic surprise solve the first-type resolution limit, i.e., the embedded communities are identified by adjusting the resolution parameter. Thirdly, we show that the multi-resolution asymptotic surprise has strong tolerance against the second-type limit[40, 42]. Fourthly, we exhibit the ability of the multi-resolution asymptotic surprise to identify the different-scale community structures in homogeneous and heterogeneous hierarchical networks.

### 3.1. Community-loop networks

In the community-loop networks, it will be more and more difficult to identify the predefined communities with the increase of $p_o/p_i$, because the difference between the inter- and intra-community link densities decreases. Some communities will be merged into one group, that is to say, the first-type resolution limit will appear. As a result, the number of identified communities (Nd) decreases and will be less than the predefined ones (see **Figure 2**(A) and (B) for two sets of test networks with $r$=8 and 64).

Normalized Mutual Information (NMI) [54] often is used to estimate the similarity between two community partitions. NMI=1 if perfectly matched; otherwise, the less the matching, the smaller the NMI. We also use the measure to estimate the amount of community information correctly obtained in the networks with known community structures. The results show that NMI is to be less than 1 with the increase of $p_o/p_i$, due to the first-type resolution limit (see **Figure 2**(C) and (D)).

As declared above, original/asymptotic surprise has higher resolution than modularity in the networks. The results indeed show that the number of identified communities as well as NMI by modularity clearly decreases more quickly, while original/asymptotic surprise can identify the communities in networks better. And the increase of network size (or the number of predefined communities) will quicken the merging of communities for all methods (see **Figure 2** for $r$= 8 and 64). Moreover, we confirm that our improved algorithms (AS-IL and OS-IL) can identify the communities in the networks more effectively than the original algorithms (AS-OL and OS-OL).

As expected, the resolution limit, or say, the merging of communities, can be solved by effective multi-resolution version of the methods. The number of predefined communities can be found correctly at suitable resolution (see **Figure 3** (A) and (B) for Nd=8 and 64). We further confirm that not only the number of predefined communities but also the predefined community structures have been identified correctly at suitable resolution (**Figure 3** (C) and (D) for NMI=1). So the multi-resolution version of the methods can help discover the embedded communities in the networks better than the original versions. Moreover, the larger the $p_o/p_i$-value, the more difficult to identify the embedded communities, because the window of resolution at which community structure can be identified decreases with the increase of $p_o/p_i$.

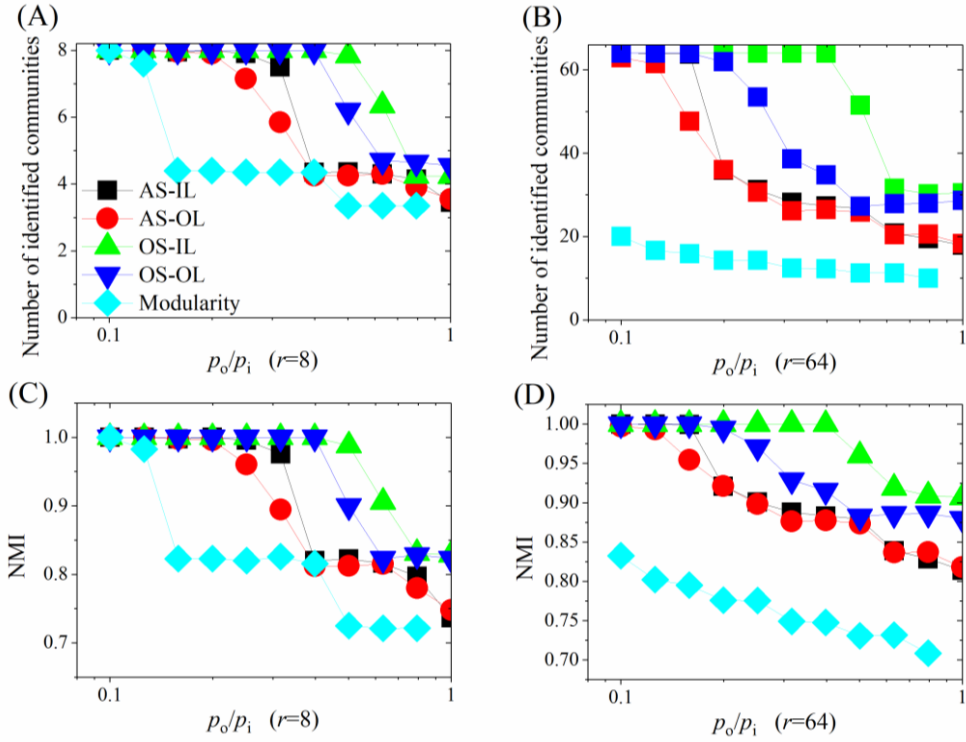

**Figure 2**. Number of identified communities by different methods, as a function of $p_o/p_i$ in the loop-community networks with (A) $r=8$ and (B) $r=64$ respectively. Normalized Mutual Information (NMI) by different methods in the networks with (C) $r=8$ and (D) $r=64$ respectively. Note that AS/OS-OL and AS/OS-IL denote asymptotic/original surprise using IL and OL algorithms respectively.

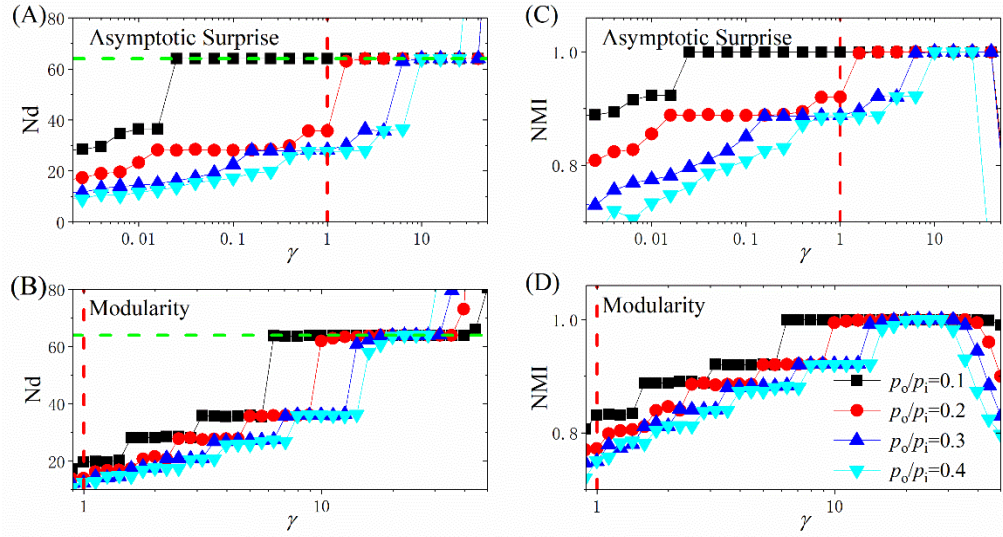

**Figure 3**. Number of identified communities (Nd) as a function of resolution parameter by different methods: (A) asymptotic surprise and (B) modularity, in the loop-community networks with $r=64$ and different $p_o/p_i$-values. Normalized Mutual Information (NMI) by (C) asymptotic surprise and (D) modularity in the networks.

## 3.2. Lancichinetti-Fortunato-Rachicchi (LFR) networks

Here, we apply the methods to a kind of networks with tunable sizes and heterogeneous structures, Lancichinetti-Fortunato-Rachicchi (LFR) networks [53], which have more realistic properties that are similar to real-word networks. In the LFR networks, the vertex degrees and community sizes are determined by the exponents of the power-law distributions $t_1$ and $t_2$ respectively; a mixing parameter $\mu$ controls the ratio between the external degree of each vertex with respect to its community and the total degree of the vertex. With the increase of $\mu$, the communities in the networks become more and more difficult to identify. Other parameters: $N$ is the number of vertices; $k_m$ and $k_{max}$ are the mean degree and maximum degrees; $C_{min}$ and $C_{max}$ are respectively the minimum and maximum community sizes. The parameters in the section are set as follows: $N$=1000, $k_m$=20, $k_{max}$=50, $C_{min}$=10 and $C_{max}$=50, t1=-2, and t2=-2.

**Figure 4**(A) shows that asymptotic/original surprise (AS/OS-IL/OL) can work very well in the LFR networks, but modularity cannot correctly identify the community structure due to the first-type resolution limit and with the increase of the mixing parameter, the resolution limit becomes more serious. We further show the fraction (Fr) of vertices affected by the merging of communities increases with the increase of the mixing parameter (see inserted graph in **Figure 4**(A)). This confirms the effect of the first-type resolution limit, especially for large mixing parameter. Moreover, we also show that our improved algorithms (AS-IL and OS-IL) can identify the communities in the networks better than the original algorithms (AS-OL and OS-OL).

Similarly to the above section, the first-type resolution limit can be solved by adjusting the resolution parameter (see **Figure 4**(B) and (C)). The pre-defined community structure can be identified correctly at suitable resolution. As we see, the smaller the mixing parameter $\mu$, the longer the length of the plateaus of $\gamma$ in logarithmic coordinate. This also means that the community structures in the networks may be more stable for the method. To a certain extent, the length of the plateaus can be regarded as a measure for stability of community structure, though it is closely related to the methods themselves. From another viewpoint, if a method has a longer plateaus of $\gamma$ than other methods in the same network, then it may find the community structure better.

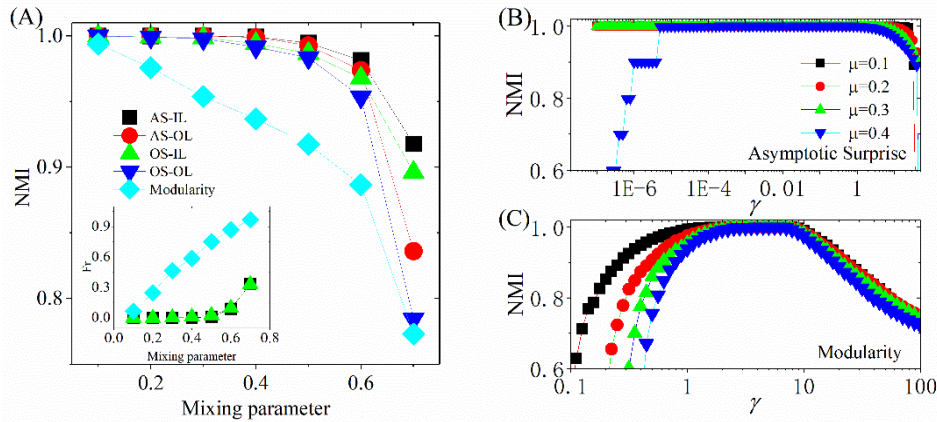

**Figure 4**. (A) Normalized mutual information (NMI) by different methods, as a function of mixing parameter in the LFR networks. Inserted graph shows the fraction (Fr) of vertices affected by the merging of communities. NMI as a function of resolution parameter by (B) asymptotic surprise and (C) modularity in the LFR networks with different μ-values. AS(OS)-OL and AS(OS)-IL denote asymptotic(original) surprise using IL and OL algorithms respectively.

**3.3. Fortunato-Barthélemy graph**

As declared in references [40, 42, 55], multi-resolution methods such as modularity can solve the first-type resolution limit of it, but still may encounter the second-type resolution limit: (large) communities may have broken up before (small) communities are identified by varying resolution parameter when the community-size difference is very large. To exhibit the second-type resolution limit and the ability of the multi-resolution asymptotic surprise against the second-type resolution limit, we apply it to the graph that consists of two large cliques with $n_1$ vertices and two small cliques with $n_2$ vertices, which was initially proposed by Fortunato and Barthélemy (FB) to show the (first-type) resolution limit of modularity [33].

**Figure 5** (A) firstly shows that the multi-resolution modularity can identify community structures of two significant scales in the network with small heterogeneity of community sizes ($n_1$=10 and $n_2$=5): the one for Nd=3 consist of two large cliques and one group with two small cliques, while another one for Nd=4 is the predefined partition. This is because the community-size difference is very small in the networks, and the second-type resolution limit does not appear for modularity.

For larger heterogeneity of community sizes, e.g., $n_1$=30 and $n_2$=5 in **Figure 5** (B), the predefined partition becomes more difficult to be identified, and the second-type limit of modularity occurs—large communities will break up before other small communities become visible. **Figure 5** (B) clearly shows, because of the second-type resolution limit, modularity cannot correctly identify the predefined community structure (NMI<1), even if by adjusting the resolution parameter. Compared to modularity, the multi-resolution asymptotic surprise can correctly do this in the two networks with small and large heterogeneity of community sizes (see **Figure 5** (C) and (D)). This indicates that the asymptotic surprise has stronger tolerance against the second-type resolution limit in the networks, though both of them have flexible resolution.

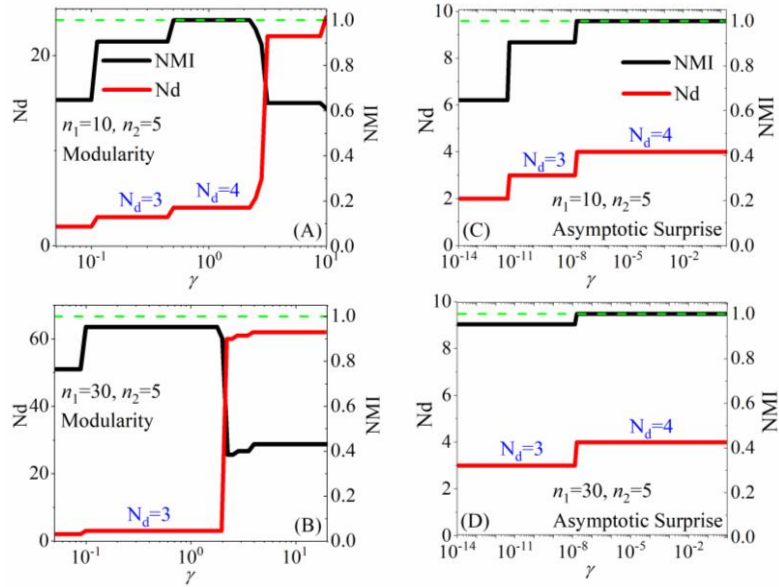

**Figure 5.** For multi-resolution modularity, the number Nd of identified communities and NMI, as a function of resolution parameter $\gamma$, in the FB networks that consist of two (large) cliques of $n_1$ vertices and two (small) cliques of $n_2$ vertices: (A) $n_1$=10 and $n_2$=5; (B) $n_1$=30 and $n_2$=5. For multi-resolution asymptotic surprise, the number Nd of identified communities and NMI, as a function of resolution parameter $\gamma$, in two FB networks with (C) $n_1$=10 and $n_2$=5; (D) $n_1$=30 and $n_2$=5.

### 3.4. Homogeneous and hierarchical network

To exhibit the ability of the multi-resolution asymptotic surprise to identify communities at different scales, we apply it to a sets of *homogeneous* and *hierarchical* networks that have 256 vertices and two predefined hierarchical levels [56]. The first level contains 16 groups of 16 vertices and the second level contains 4 groups of 64 vertices. The number of links of each vertex with the most internal community is $k_{in0}$, the number of links of each vertex with the most external community is $k_{in1}$, and the number of links with any other vertex at random in the network is 1.

**Figure 6** shows that the multi-resolution version of modularity and Significance can identify the community structures at two scales, which are marked respectively by L1 and L2. Moreover, the decrease of $k_{in1}$ leads to the decrease of the needed $\gamma$-value for the identification of L1-level communities, because it leads to the decrease of the number of links between L1-level communities and thus L1-level communities are more easily to be disconnected.

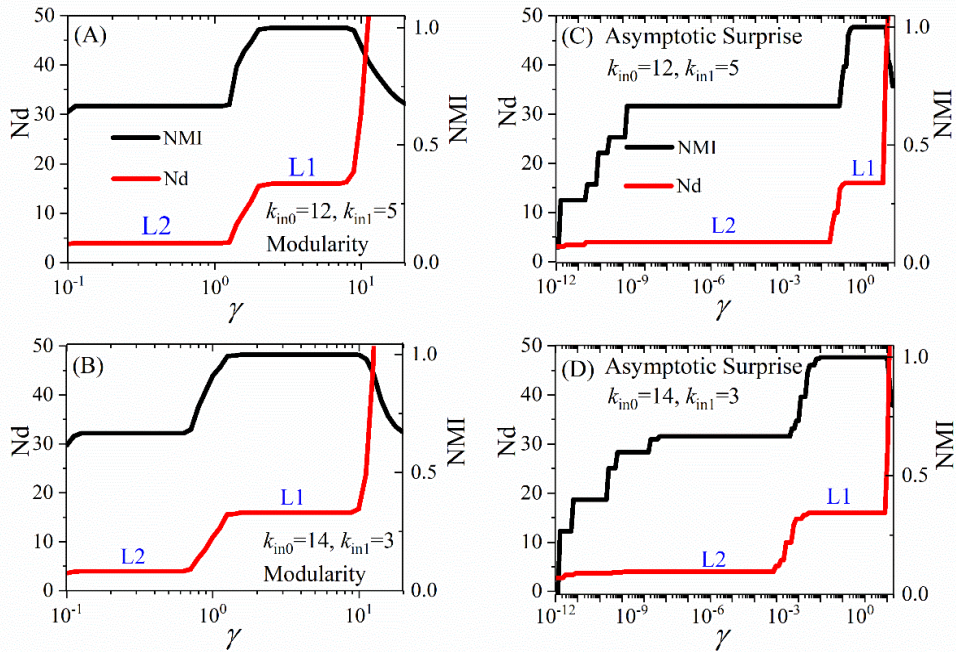

**Figure 6.** The number Nd of identified communities and NMI as a function of resolution parameter $\gamma$, by different methods, in the homogeneous and hierarchical networks with two-scale community structures. For multi-resolution modularity, (A) $k_{in0}=12$ and $k_{in1}=5$; (B) $k_{in0}=14$ and $k_{in1}=3$. For multi-resolution asymptotic surprise, (C) $k_{in0}=12$ and $k_{in1}=5$; (D) $k_{in0}=14$ and $k_{in1}=3$. Note that L1 and L2 highlight two predefined scales in the networks, which are correctly identified.

### 3.5. Heterogeneous and hierarchical networks

Then, the multi-resolution asymptotic surprise is applied to a sets of *heterogeneous* and *hierarchical* networks with two scales [57]. In the networks, the number of vertices is 1000; the average degree is 20; maximum degree is 50; minimum and maximum for micro community sizes are 10 and 25; minimum and maximum for macro community sizes are 50 and 100; $\mu_1$ and $\mu_2$ control the mixing parameters for the macro and micro communities; other parameters are default[1].

**Figure 7**(A) and (B) show that modularity with $\gamma=1$ can only identify the macro

---
[1] https://sites.google.com/site/santofortunato/inthepress2

community structure (marked by L2), while the multi-resolution modularity can identify the community structures at the micro and macro scales marked by L1 and L2. Similarly, **Figure 7**(C) and (D) show that the asymptotic surprise with $\gamma=1$ can only identify the micro community structure (marked by L1), while the multi-resolution asymptotic surprise can correctly identify the community structures at the two scales marked by L1 and L2. Moreover, the larger $\gamma$-value is needed for the identification of macro communities with the increase of $\mu_1$, because this leads to the increase of the number of links between macro communities.

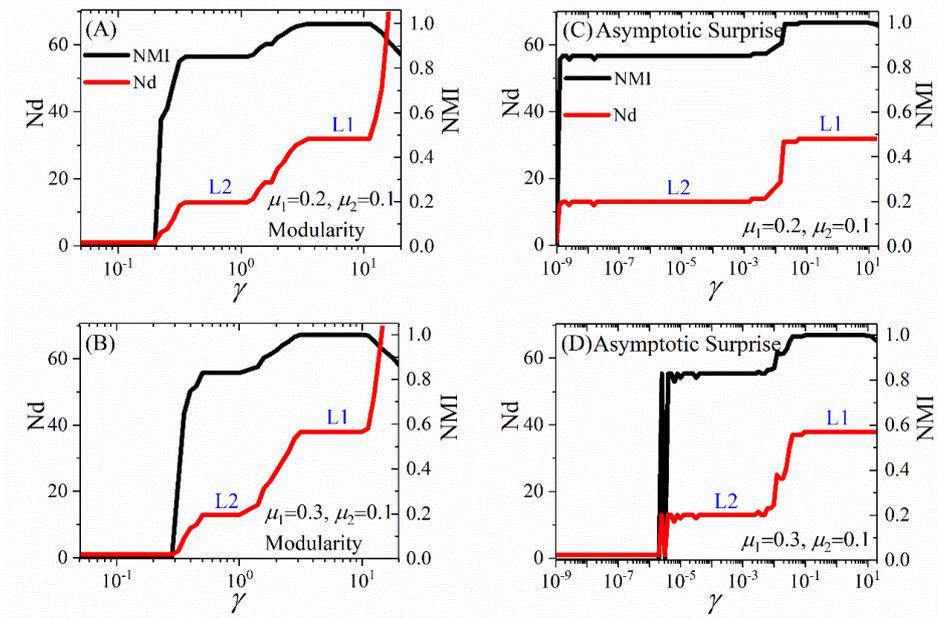

**Figure 7.** The number Nd of identified communities and NMI as a function of resolution parameter $\gamma$ by different methods, in the heterogeneous hierarchical networks with two-scale community structures. For multi-resolution modularity: (A) $\mu_1=0.2$ and $\mu_2=0.1$; (B) $\mu_1=0.3$ and $\mu_2=0.1$. For multi-resolution asymptotic surprise: (C) $\mu_1=0.2$ and $\mu_2=0.1$; (D) $\mu_1=0.3$ and $\mu_2=0.1$. Note that L1 and L2 highlight two predefined scales in the networks.

## 4. Conclusion

Community structure is an important topological property of complex networks. Many methods have been proposed to identify the community structure in complex networks, and optimizing statistical measures for community structures is one of most popular strategies for community detection, such as modularity, Hamiltonians, surprise as well as asymptotic surprise. On the one hand, understanding the (critical) behaviors of the methods is necessary, because each of them has respective scope of application. On the other hand, some of the methods lack the flexibility of resolution. This is incompatible with multi-scale communities of networks.

Here, we discussed the phase transition of asymptotic surprise in community detection. The asymptotic surprise generally has higher resolution than modularity, but there still exists the resolution limit, which is closely related to such network parameters as the intra- and inter-link densities. According to the theoretical analysis of the resolution limit, a multi-resolution method based on asymptotic surprise was introduced, which is a generalization of asymptotic surprise to multi-scale networks. Moreover, to optimize asymptotic surprise more effectively, we proposed an improved Louvain algorithm by using an effective initialization process and a refining process.

By a series of experimental tests in various networks, we firstly displayed the first-type resolution limit of the asymptotic surprise as well as the effectiveness of our improved Louvain algorithm. By the resolution parameter, the multi-resolution asymptotic surprise can solve its (first-type) resolution limit. Then, we showed the second-type resolution limit for multi-resolution methods—(large) communities may break up before (small) communities become visible when community-size difference is very large. The results showed that, for large heterogeneity of community sizes, the multi-resolution modularity is easily to encounter the second-type limit, while the multi-resolution asymptotic surprise can do well in the networks, because it has stronger tolerance against the second-type resolution limit in the networks. Finally, we validated the effectiveness of the multi-resolution asymptotic surprise in discovering the multi-scale communities in the hierarchical networks, including a set of homogeneous networks and a set of heterogeneous networks.

Overall, the extension of asymptotic surprise to multi-scale networks provides an alternative approach to study multi-scale networks, while there might be other extension of asymptotic surprise in the future. We expect that this work could help further understand the asymptotic surprise in community detection and provide useful insight into the study of community structure in complex networks.

**Acknowledgement**

This work was supported by the construct program of the key discipline in Hunan province, the Training Program for Excellent Innovative Youth of Changsha, the National Natural Science Foundation of China (Grant No. 61702054, 71871233, 61622213 and 61832019), the Hunan Provincial Natural Science Foundation of China (Grant No. 2018JJ3568), the Scientific Research Fund of Education Department of Hunan Province (Grant No. 17A024), the Scientific Research Project of Hunan Provincial Health and Family Planning Commission of China (Grant No. C2017013), and the Beijing Natural Science Foundation (Grant No. 9182015).

**Appendix**

To display the critical behaviors of statistical measures for community structures, we introduced a set of community-loop networks where $r$ communities are connected one by one. In the networks, each community has $n_c$ vertices, and the whole network has $n = r \cdot n_c$ vertices. $p_i$ denotes the probability of linking vertices within community; $p_o$ denotes the probability of linking vertices respectively in two distinct and adjacent communities. Consider a set of partitions with $r/x$ groups of vertices, where each group has $x$ adjacent communities. For the partitions,

$$\begin{aligned} m_{in} &= r \cdot n_c^2 p_i + \frac{2r}{x}(x-1) n_c^2 p_o \\ m &= r \cdot n_c^2 p_i + 2r \cdot n_c^2 p_o \\ M_{in} &= \frac{r}{x}(xn_c)^2 \\ M &= (r \cdot n_c)^2 \end{aligned} \qquad (8)$$

while the probability of a link existing within a community and its expected value can be written as,

$$q_x = \frac{1 + \frac{2p_o}{p_i}\frac{(x-1)}{x}}{1 + \frac{2p_o}{p_i}} = \frac{1 + 2\xi \frac{(x-1)}{x}}{1 + 2\xi} = 1 - \frac{2\xi}{1+2\xi}\frac{1}{x}, \quad (9)$$

and,

$$\bar{q}_x = \frac{x}{r}, \quad (10)$$

where $\xi = p_o/p_i$. As a result, the asymptotic surprise, as a multivariate function, can be written as,

$$\begin{aligned} S_x &= mD(q_x \| \bar{q}_x) \\ &= m\left(q_x \ln\left(\frac{q_x}{\bar{q}_x}\right) + (1-q_x)\ln\left(\frac{1-q_x}{1-\bar{q}_x}\right)\right) \\ &= m\left(\left(1 - \frac{2\xi}{1+2\xi}\frac{1}{x}\right)\ln\left(\left(1 - \frac{2\xi}{1+2\xi}\frac{1}{x}\right)\bigg/\frac{x}{r}\right) + \frac{2\xi}{1+2\xi}\frac{1}{x}\ln\left(\frac{2\xi}{1+2\xi}\frac{1}{x}\bigg/\left(1-\frac{x}{r}\right)\right)\right) \end{aligned} \quad (11)$$

For partition x=1,

$$S_1 = m\left(q_1 \ln\left(q_1 \bigg/ \frac{1}{r}\right) + (1-q_1)\ln\left((1-q_1)\bigg/\left(1-\frac{1}{r}\right)\right)\right). \quad (12)$$

For partition x=2,

$$S_2 = m\left(q_2 \ln\left(q_2 \bigg/ \frac{2}{r}\right) + (1-q_2)\ln\left((1-q_2)\bigg/\left(1-\frac{2}{r}\right)\right)\right). \quad (13)$$

Communities will merge if $\Delta S = (S_2 - S_1)/m > 0$. By using $r-1 \approx r-2 \approx r$ for large $r$-value,

$$\begin{aligned} \Delta S &\approx q_1 \ln\left(\frac{q_2}{2q_1}\right) + (1-q_1)\ln\left(\frac{1-q_2}{1-q_1}\right) + \beta \ln\left(\frac{q_2}{1-q_2}\frac{r}{2}\right) \\ &\approx -D(q_1 \| q_2) - q_{x_1}\ln(2) + \beta \ln\left(\frac{q_2}{1-q_2}\frac{r}{2}\right) \end{aligned} \quad (14)$$

where $\beta = \frac{p_o/p_i}{1+2p_o/p_i} = \frac{\xi}{1+2\xi}$. By solving $\Delta S = 0$ for $r$, the critical number of communities is obtained,

$$\begin{aligned} r^* &\approx \frac{1-q_2}{q_2}\exp\left(\frac{1}{\beta}D(q_1 \| q_2)\right)x^{\frac{q_1}{\beta}+1} \\ &= \frac{p_o/p_i}{1+p_o/p_i}\left(\frac{1}{1+p_o/p_i}\right)^{\frac{p_i}{p_o}} 2^{\frac{p_i}{p_o}+3} \\ &= \frac{\xi}{1+\xi}\left(\frac{1}{1+\xi}\right)^{\frac{1}{\xi}} 2^{\frac{1}{\xi}+3} \end{aligned} \quad (15)$$